\documentclass[]{JHEP3}

\usepackage{epsfig}
\usepackage{amsmath,amssymb}
\usepackage{citesort}

\allowdisplaybreaks[1]
\newcommand{\nn}{\nonumber}
\newcommand{\dsl}{\hspace{-5pt}/}
\newcommand{\aki}{\hspace*{8pt}}

\newcommand{\im}{{\rm Im}}

\newcommand{\TL}[1]{T^{\ma}_{L#1}}
\newcommand{\TR}[1]{T^{\ma}_{R#1}}
\newcommand{\ma}{\mu\alpha}

\title{$T$-odd asymmetries in radiative top-quark decays}

\author{Kaoru Hagiwara\\
        KEK Theory Division and Sokendai, Tsukuba 305-0801, Japan\\
	E-mail: \email{kaoru.hagiwara@kek.jp}}
\author{Kentarou Mawatari\\
        School of Physics, Korea Institute for Advanced Study,
        Seoul 130-722, Korea\\
        E-mail: \email{kentarou@kias.re.kr}}
	\author{Hiroshi Yokoya\footnote{
	Address from December 1, 2007: Theory Unit, Physics Department,
	CERN, CH-1211 Geneva 23, Switzerland; E-mail:
        hiroshi.yokoya@cern.ch}\\
        Department of Physics, Niigata University,
        Niigata 950-2181, Japan\\
        E-mail: \email{yokoya@nt.sc.niigata-u.ac.jp}}

\abstract{
We study the angular distribution of the charged lepton in the
 top-quark decay into a bottom quark and a $W$ boson which subsequently
 decays into $\ell\nu_{\ell}$, when a hard gluon is radiated off.
The absorptive part of the $t\to bWg$ decay amplitudes, which gives rise
 to $T$-odd asymmetries in the distribution, is calculated at the
 one-loop level in perturbative QCD.
The asymmetries at a few percent level are predicted, which may be
 observable at future colliders.
}

\preprint{KEK-TH-1140\\ KIAS-P07005 \\ \today}

\begin{document}

\section{Introduction}

$T$-odd effects in hard QCD processes have been attracting our
 attentions for more than 30 years, but no experimental verification of
 the predictions~\cite{DeRujula:1978bz,Korner:1980np,Hagiwara:1981qn,%
Hagiwara:1982cq,Pire:1983tv,Hagiwara:1984hi,Hagiwara:1991xy,%
Hagiwara:1990dx,Brandenburg:1995nv}
 has been presented yet.
$T$-odd observables change sign under the operation of reversing both
 the spatial momenta and the spins of the all the particles without
 interchanging initial and final states; see
 Refs.~\cite{DeRujula:1972te,Hagiwara:1982cq} for details.%
\footnote{$T$-odd effects are sometimes referred to as
 na\"ive-$T$-odd~\cite{Hagiwara:2006qe} or
 $T_N$-odd~\cite{Brandenburg:1995nv} in order to distinguish them from
 the genuine time-reversal operation $T$, which exchanges the initial and
 the final states.
}
In $T$-invariant theories like perturbative QCD, the $T$-odd effects
 arise due to the re-scattering phase, or the absorptive part of the
 amplitudes, which appears in the loop level.
Such $T$-odd quantities in hard processes can be predicted in
 perturbative QCD, and should be tested experimentally.

Since de R\'ujula et al.\ proposed to measure $T$-odd effects as an
 experimental test of the non-abelian nature of QCD in
 $e^+e^-\to\Upsilon\to ggg$ with a longitudinally polarized
 beam~\cite{DeRujula:1978bz}, several theoretical studies have been
 performed for the quark and gluon processes with an electroweak
 current.
They can be classified into three types:
\begin{enumerate}
\item
 Three jets in $e^+e^-$ annihilation with a longitudinally polarized
     beam, $e^+e^-\to q\bar qg$~\cite{Korner:1980np,Hagiwara:1990dx,%
Brandenburg:1995nv}.
\item
 Semi-inclusive deep-inelastic neutrino~\cite{Hagiwara:1981qn} or
  longitudinally polarized electron~\cite{Hagiwara:1982cq,Ahmed:1999ix}
  scattering, $\ell p \to \ell' h X$.
\item
 Drell-Yan-type process, $p\bar p\to\gamma^*/W/Z+{\rm jet}+X$.
 References~\cite{Pire:1983tv,Carlitz:1992fv,Yokoya:2007xe} considered
  single-spin asymmetry in the Drell-Yan process with longitudinally
  polarized proton beam, while $T$-odd effects without spin
  measurement were studied in
  $W$-jet~\cite{Hagiwara:1984hi,Hagiwara:2006qe} and
  $Z$-jet~\cite{Hagiwara:1991xy} events at hadron colliders.
\end{enumerate}
The absorptive parts of the relevant one-loop amplitudes in these three
 processes are related to each other through
 crossing~\cite{Korner:2000zr}.
In addition to above three processes, there also exists another $T$-odd
 observable, the normal polariation in top-quark pair-production at
 $e^+e^-$ and hadron colliders~\cite{Kuhn:1985ps,Kane:1991bg,%
Bernreuther:1992ef,Bernreuther:1995cx,Dharmaratna:1996xd,%
Brandenburg:1998xw}.

Observations of $T$-odd effects in hard processes are a challenging task
 since they do not appear at the tree level.
So far, no experimental test has been made for the above
 processes~\cite{Abe:1995ge,Airapetian:1999tv,Chekanov:2006gt}, even
 though large non-perturbative $T$-odd effects have been observed in
 hadron spin physics~\cite{Airapetian:1999tv,pnlambda}.
We may note that the possibility to observe the perturbative
 $T$-odd effects in $W$-jet events at the Tevatron run II has recently
 been pointed out in~\cite{Hagiwara:2006qe}.\\

In this article, we propose a new measurement of the $T$-odd effects in
 radiative top-quark decays.
We study $T$-odd angular distributions of $W$-decay leptons in the
 radiative top-quark decay into a bottom quark, a $W$ boson, and a
 gluon:
\begin{align}
 t \to b + W^+ + g; \quad W^+ \to \ell^+ + \nu_\ell. \label{process}
\end{align}
Due to the large mass, $m_t=175$ GeV, top-quark decay is not affected by
 hadronization, and hence it can be dictated by perturbative QCD.
Even though the correction up to ${\cal O}(\alpha_s^2)$ to the total
 decay width of the top quark is known~\cite{Czarnecki:1998qc}, the
 correction to the $W$-decay lepton angular distribution in the top
 quark decay has been calculated only up to ${\cal
 O}(\alpha_s)$~\cite{Fischer:2001gp}.
We calculate the absorptive part of the amplitudes for the $t\to bWg$
 process in the one-loop order ${\cal O}(\alpha_s^2)$, which gives the
 leading contribution to the $T$-odd asymmetries.
The predictions may be tested at future colliders such as the Large
 Hadron Collider (LHC) and the International Linear Collider (ILC).\\

The article is organized as follows.
In Sec.~\ref{sec:form}, we present the lepton decay distribution using
 the density matrices of the $t\to bWg$ decay and the $W\to \ell\nu$
 decay, and give the general kinematics relevant to our analysis.
In Sec.~\ref{sec:dis}, after showing the $T$-even lepton angular
 distributions, we discuss the $T$-odd distributions in detail, and study
 their observability at future experiments.
In Sec.~\ref{sec:spin}, we consider radiative decays of polarized
 top-quarks and discuss another $T$-odd observable, the angular
 correlation between the top-quark spin and the decay plane.
Section~\ref{sec:sum} is devoted to a summary.
In appendix~\ref{app:amp}, we give the absorptive part of the $t\to bWg$
 decay amplitudes in the one-loop order by using the Feynman parameter
 integral calculation.
In appendix~\ref{app:loop}, we present our results in terms of the loop
 scalar functions.

\section{$t\to bW^+g$ decay density matrix}\label{sec:form}

The decay rate of the process (\ref{process}) can be
 expressed in terms of the $t\to bWg$ decay and the $W\to\ell\nu$ decay
 density matrices in the narrow width approximation of the $W$ boson,
\begin{align}
 d\Gamma=\sum_{\lambda,\lambda'}d\Gamma_{\lambda\lambda'}^{t}\,
 \frac{1}{\Gamma_W}\, d\Gamma_{\lambda\lambda'}^{W}, \label{doubleDM}
\end{align}
where $\Gamma_W$ is the total decay width of $W$ boson, and $\lambda$,
 $\lambda'=\pm$, 0 denote the $W$-boson helicity.
The $3\times 3$ $W$-polarization density matrix for the $W^+$ decay
 reads
\begin{align}
  \frac{1}{\Gamma_W}\frac{d\Gamma_{\lambda\lambda'}^{W}}{d\cos\theta\,
 d\phi} =
 B_{\ell}\,\frac{3}{8\pi}\,L_{\lambda\lambda'}(\theta,\phi)\label{wDM}
\end{align}
with the decay branching fraction $B_{\ell}=B(W\to\ell\nu)$ and
\begin{align}
 L_{\lambda\lambda'}(\theta,\phi) =
  \begin{pmatrix}
   \frac{(1+\cos\theta)^2}{2}
  &\frac{\sin\theta(1+\cos\theta)}{\sqrt{2}}e^{i\phi}
  &\frac{\sin^2\!\theta}{2}e^{2i\phi}
  \\
   \frac{\sin\theta(1+\cos\theta)}{\sqrt{2}}e^{-i\phi}
  &\sin^2\theta
  &\frac{\sin\theta(1-\cos\theta)}{\sqrt{2}}e^{i\phi}
  \\
   \frac{\sin^2\!\theta}{2}e^{-2i\phi}
  &\frac{\sin\theta(1-\cos\theta)}{\sqrt{2}}e^{-i\phi}
  &\frac{(1-\cos\theta)^2}{2}
 \end{pmatrix}.
\end{align}
Here, the $3\times 3$ matrices are for $\lambda$, $\lambda'=(+,0,-)$,
 and the polar and azimuthal angles $(\theta,\phi)$ of the charged
 lepton are defined in the rest frame of the $W$ boson, where the
 $z$-axis is taken along the $W$ momentum direction in the rest frame of
 the top quark.
The $x$-axis ($\theta=\pi/2$, $\phi=0$) is in the $t\to bWg$ decay plane
 as explained below.\\

Before we show the $t\to bWg$ density matrix
 $d\Gamma^{t}_{\lambda\lambda'}$, we define the kinematical variables
 for the process
\begin{align}
 t(p_t,\sigma_t) \to
 b(p_b,\sigma_b) + W^+(q,\lambda) + g(p_g,\sigma_g),
\end{align}
 where the four-momenta of each particle are defined in the top rest
 frame as
\EPSFIGURE[r]{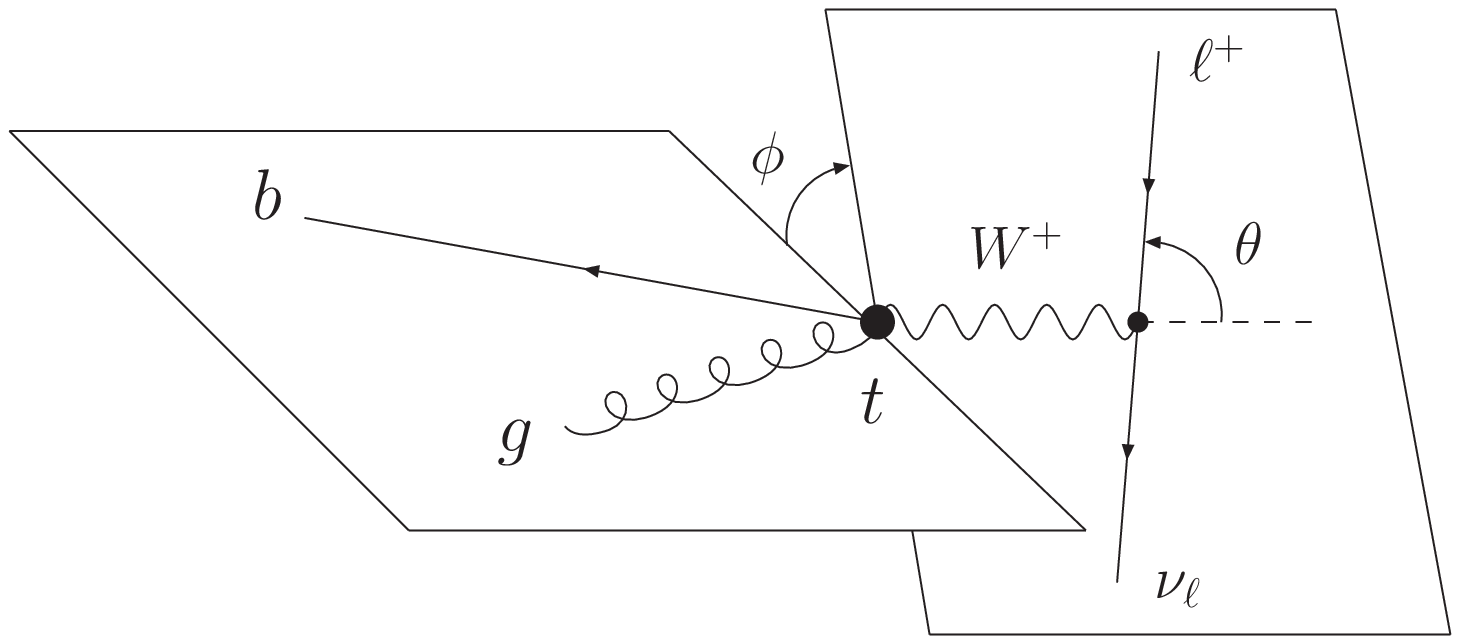,width=.5\textwidth,clip}
 {Schematic view of the coordinate system for the $t\to bW^+g$ decay,
 followed by the $W^+\to \ell^+\nu_{\ell}$ decay.\label{frame}}
\begin{align}
 & p_t^{\mu} = (m_t, 0, 0, 0), \nn\\
 & p_b^{\mu} = (E_b,p_b\sin\hat\theta,0,p_b\cos\hat\theta), \nn\\
 & q^{\mu}   = (E_W, 0, 0, q), \nn\\
 & p_g^{\mu} = (E_g,p_{g,x},0,p_{g,z}).\label{fourmomenta}
\end{align}
Helicities of each particle, $\sigma_t$, $\sigma_b$, $\lambda$ and
 $\sigma_g$, are also defined in the top rest frame.
The $z$-axis is taken along the $W$ boson momentum, and $y$-axis is
 along $\vec q\times\vec p_b$, the normal of the decay plane; see
 Fig.~\ref{frame}.

We define the dimensionless variables as
\begin{align}
 (z_1, z_2, z_3)&\equiv\left( \frac{2p_t\!\cdot\!p_b}{m_t^2},
 \frac{2p_t\!\cdot\!q}{m_t^2},\frac{2p_t\!\cdot\!p_g}{m_t^2}\right)
=\left(\frac{2E_b}{m_t}, \frac{2E_W}{m_t}, \frac{2E_g}{m_t}\right).
\end{align}
These are the energy fraction of $b$, $W$ and $g$, respectively, and
 satisfy the energy conservation condition, $z_1+z_2+z_3=2$.
The kinematically allowed region is given in the $(z_1,z_2)$ plane by
\begin{align}
 &2y\le z_1 \le 1-x^2+y^2, \quad 2x\le z_2 \le 1+x^2-y^2,\nn\\
 &(z_1^2-4y^2)(z_2^2-4x^2)
 -\left[2+2x^2+2y^2-2z_1-2z_2+z_1 z_2\right]^2\ge 0,\label{bound}
\end{align}
 with $x=m_W/m_t$ and $y=m_b/m_t$.

The mass of the $b$-quark is kept to be finite ($m_b=4$ GeV) for the
 tree-level calculation.
However, as we will see later, the effect of the mass is negligible.
Thus, for the calculation of the $T$-odd distributions, we take the
 $m_b=0$ limit, which simplifies the framework of the one-loop
 calculation.
In the case that we ignore the $b$-quark mass, there appears a
 kinematical singularity in the $z_2\to 1+x^2$ limit, when the $b$-quark
 and gluon momenta are collinear.
An infra-red (IR) singularity also exists at $z_3\to 0$, where the
 emitted gluon becomes soft.\\

Let us now present the density matrix for the $t\to bWg$ decay,
 $d\Gamma_{\lambda\lambda'}^{t}$ in Eq.~(\ref{doubleDM}).
The matrix elements of the $t\to bWg$ decay are expressed as
\begin{align}
 i {\cal M}_\lambda = \frac{-igg_s}{\sqrt{2}}t^{a}V_{tb}\;
    \bar{u}(p_b,\sigma_b)\;T^{\ma}\,u(p_t,\sigma_t)
    \ \epsilon^{*}_{\mu}(q,\lambda)\,
   \epsilon^{a*}_{\alpha}(p_g,\sigma_g), \label{amp}
\end{align}
 where $g$ and $g_s$ are the weak and strong coupling constants,
 $t^a$ is the SU(3) color matrix, and $V_{tb}$ is the
Cabibbo-Kobayashi-Maskawa (CKM) matrix element.
The tensor $T^{\ma}$ is a $4\times 4$ matrix in the spinor space.
The leading contribution to the real part of $T^{\ma}$ comes from
 the tree diagrams \cite{Couture:1989ru,Mrenna:1991wd}, while the
 imaginary part appears first in the one-loop diagrams.
All the tree and the one-loop diagrams needed in our analysis are shown
 in Fig.~\ref{feynman}.
We give details of our calculation of $T^{\ma}$ in the
 appendices.

\EPSFIGURE[t]{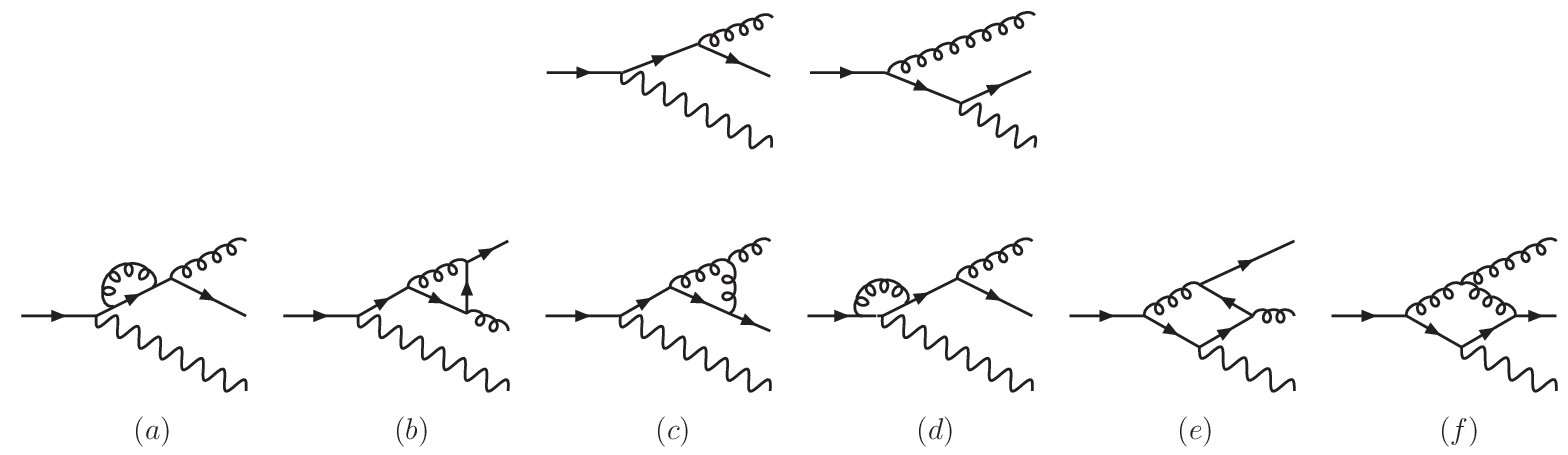,width=1\textwidth,clip}
{Feynman diagrams for the $t\to bWg$ decay~\cite{Binosi:2003yf}.
The top two are the tree level diagrams, and the bottom six are the
 one-loop level diagrams contributing to the absorptive part of the
 amplitudes.\label{feynman}}

Factorizing the color factor and the coupling constants, we define
 the reduced density matrix $H_{\lambda\lambda'}$ as
\begin{align}
\overline{\sum} {\cal M}_{\lambda}{\cal M}^{*}_{\lambda'} =
 4\sqrt{2}\pi G_{F} \alpha_{s} m_W^2 |V_{tb}|^2 C_F \cdot
 H_{\lambda\lambda'}.
\end{align}
The summation stands for the sum/average of the spins of the particles
 except $W$ boson and the sum/average of colors.
In terms of $H_{\lambda\lambda'}$, the density matrix
 $d\Gamma_{\lambda\lambda'}^{t}$ is expressed as
\begin{align}
 \frac{d\Gamma^{t}_{\lambda\lambda'}}{dz_1 dz_2}
 =\frac{G_F \alpha_{s} m_t^3 x^2 |V_{tb}|^2 C_F}{32\sqrt{2}\pi^2}\,
  H_{\lambda\lambda'}(z_1,z_2).\label{tDM}
\end{align}
Finally, combining the top- and $W$-decay density matrices in
 Eqs.~(\ref{tDM}) and (\ref{wDM}), the decay distribution in
 Eq.~(\ref{doubleDM}) is expressed as
\begin{align}
&\frac{d\Gamma}{dz_1 dz_2\, d\cos\theta\, d\phi}=
\frac{3B_{\ell} G_F \alpha_{s}m_t^3 x^2 |V_{tb}|^2 C_F}
{256\sqrt{2}\pi^3}
\sum_{\lambda,\lambda'}H_{\lambda\lambda'}(z_1,z_2)\,
L_{\lambda\lambda'}(\theta,\phi)\nn\\
&\qquad\equiv K\big[\,
  F_{1}(1+\cos^2\theta) + F_{2}(1-3\cos^2\theta)
+ F_{3}\sin2\theta\cos\phi + F_{4}\sin^2\theta\cos2\phi \nn \\
&\qquad+F_{5}\cos\theta + F_{6}\sin\theta\cos\phi
+ F_{7}\sin\theta\sin\phi + F_{8}\sin2\theta\sin\phi
+ F_9\sin^2\theta\sin2\phi\,\big],\label{leptondis}
\end{align}
 where $K$ is the factor in front of the summation symbol in the first
 line.
The nine independent functions $F_{1-9}(z_1,z_2)$ are defined in terms
 of the reduced density matrices $H_{\lambda\lambda'}$ as
\begin{align}
 &  F_{1} = \frac{1}{2}\left(H_{++}+H_{00}+H_{--}\right),
 && F_{6} = \frac{1}{\sqrt{2}}
\left(H_{+0}+H_{0+}+H_{-0}+H_{0-}\right),\nn\\
 &  F_{2} = \frac{1}{2}H_{00},
 && F_{7} = \frac{i}{\sqrt{2}}
\left(H_{+0}-H_{0+}-H_{-0}+H_{0-}\right),\nn\\
 &  F_{3} = \frac{1}{2\sqrt{2}}
\left(H_{+0}+H_{0+}-H_{-0}-H_{0-}\right),
 && F_{8} = \frac{i}{2\sqrt{2}}
\left(H_{+0}-H_{0+}+H_{-0}-H_{0-}\right),\nn\\
 &  F_{4} = \frac{1}{2}\left(H_{+-}+H_{-+}\right),
 && F_{9} = \frac{i}{2}\left(H_{+-}-H_{-+}\right). \nn\\
 &  F_{5} = H_{++} - H_{--}, \label{F1to9}
\end{align}
The terms independent of the azimuthal angle, $F_{1}$, $F_{2}$ and
 $F_{5}$, are provided from the diagonal terms of the density
 matrix, while the azimuthal-angle dependent terms are provided
 from the off-diagonal terms, i.e.\ the interference between the
 different polarization states of the $W$ boson.
The terms $F_1$ through $F_6$ are $T$-even, and the leading contribution
 comes from the tree diagrams.
On the other hand, $F_7$ to $F_9$ are $T$-odd, and they receive the
 leading contribution from the absorptive part of the one-loop
 amplitudes through the interference with the tree amplitudes.
Parity transformation changes the sign of $\phi$, thus $F_{7,8,9}$ are
 not only $T$-odd but also parity-odd ($P$-odd).
Assuming $CP$ invariance, the lepton angular distribution for the
 anti-top-quark decay, $\bar{t}\to \bar{b}W^-g;\
 W^-\to\ell^-\bar{\nu}_{\ell}$, can be obtained by changing the sign of
 $F_{7,8,9}$ in Eq.~(\ref{leptondis}).

\section{Lepton decay distributions}\label{sec:dis}

In this section, we present numerical results for the $T$-even and
 $T$-odd lepton angular distributions in radiative top-quark decays.
Note that, in our leading-order analysis, the $T$-even distributions
 $F_{1-6}$ are ${\cal O}(\alpha_s)$, while $T$-odd distributions
 $F_{7,8,9}$ are ${\cal O}(\alpha_s^2)$.

\subsection{$T$-even distributions}

In Fig.~\ref{contour}, we show a contour plot of the function
 $F_1(z_1,z_2)$, which gives the total rate of the $t\to bWg$ decay,
 $d\Gamma/dz_1dz_2=K\,(16\pi/3)\,F_1$, after integrating over the lepton
 decay angles.
The kinematical boundary given by Eq.~(\ref{bound}) for $m_t=175$~GeV,
 $m_W=80.4$~GeV and $m_b=4$ GeV ($m_b=0$) is shown by the thick (thin)
 dotted line.
To avoid the IR region near $z_2={z_2}_{\rm max}\sim 1.2$, we impose the
 $k_T$ cut,
\begin{align}
 k_T^2 \equiv 2 \min(p_b^2,p_g^2)\,(1-\cos\theta_{bg})
       >(20\,{\rm GeV})^2,\label{kt}
\end{align}
\EPSFIGURE[r]{contour,width=.48\textwidth,clip}
{Contour plot of $F_1(z_1,z_2)$ in the tree level.
$z_1$ and $z_2$ are the energy fraction of the bottom quark and the $W$
 boson, respectively.
The dotted line denotes the kinematical boundary; the dashed and
 dot-dashed lines are for the kinematical cuts for $k_T>20$ GeV and
 $\cos\theta_{bg}>-0.9$, respectively.
The thick contours are obtained for $m_b=4$ GeV, whereas the thin
 contours are for $m_b=0$.\label{contour}}

\noindent
 where $\theta_{bg}$ is the angle between the $b$-quark and gluon
 momenta in the top rest frame, shown by the dashed line. Furthermore,
 we apply the following cut:
\begin{align}
 \cos\theta_{bg}>-0.9, \label{cosbg}
\end{align}
 shown by the dot-dashed line, in order to avoid the configuration where
 the $b$-quark and gluon jets are anti-collinear.
These two cuts enable us to define the top decay plane spanned by $\vec
 p_b$ and $\vec p_g$, from which the azimuthal angle $\phi$ of the decay
 lepton is measured (see Fig.~\ref{frame}).

The decay rate is large in the region where $z_2$ is large, because of
 the collinear singularity in the $m_b=0$ limit.
As the figure shows, the effect of the $b$-quark mass is small for the
 decay-rate itself, however the kinematical boundary as well as the
 cuts are changed slightly by the mass.
\\

Next, we define the differential asymmetries as
\begin{align}
 A_i(z_2)\equiv\int dz_1\, F_i(z_1,z_2)\Big/\int dz_1\, F_1(z_1,z_2)
 \label{asym}
\end{align}
 for $i=2$ to 9.
In Fig.~\ref{z2dis_teven}, we show the $z_2$ distributions of the
 $T$-even asymmetries $A_{2,\cdots,6}$ at the tree level for the three
 $z_1$ regions: ${z_1}_{\rm min}<z_1<0.4$, $0.4<z_1<0.55$ and
 $0.55<z_1<{z_1}_{\rm max}$, with the same kinematical cuts as in
 Fig.~\ref{contour}.
The $z_2$ distributions of $F_1$ for the same $z_1$ regions are also
 shown as a reference.
In all the figures, predictions for $m_b=4$ GeV and the massless
 $b$-quark limit are shown by thick and thin lines, respectively.
Except for $F_1$, the lines for $m_b=4$ GeV and those for $m_b=0$ are
 almost degenerated.
Small difference in $F_1$ at large $z_2$ and small $z_1$ arises because
 of the difference in the kinematical boundary, as shown in
 Fig.~\ref{contour}.
\FIGURE[t]{\hspace*{4.5pt}
\epsfig{file=z2dis_teven1.eps,width=.985\textwidth,clip}
\epsfig{file=z2dis_teven2.eps,width=1.\textwidth,clip}
\caption{The $z_2$ distributions of the $T$-even asymmetries $A_{2}$ to
 $A_{6}$ at the tree level. Three cases for the different $z_1$
 regions with the same kinematical cuts as Fig.~\ref{contour} are
 shown.
Thick lines are for $m_b=4$ GeV, and thin lines are for $m_b=0$.
The distributions of $F_1$ integrated for $z_1$ are also shown as a
 reference.\label{z2dis_teven}}}

The asymmetries in the polar angular distribution $A_{2,5}$ are
 predicted to be large, more than the azimuthal angular asymmetries
 $A_{3,4,6}$.
When the $W$-boson energy (i.e., $z_2$) is large, the kinematics of the
 $t\to bWg$ three-body decays becomes close to that of the $t\to bW$
 two-body decays.
Near $z_2={z_2}_{\rm max}$, this leads to the well known results:
(i) The asymmetry $A_2$, which dictates the fraction of the decay to the
 longitudinally polarized $W$ bosons, reaches 0.7.
(ii) The fraction to the left-handed $W$ bosons is $\sim$ 0.3, and the
 fraction to the right-handed $W$ bosons is negligible.
This corresponds to the asymmetry
 $A_5\propto H_{++}-H_{--}\sim -H_{--}$.
The difference of the factor 2 comes from the normalization in
 Eq.~(\ref{F1to9}).
(iii) The $A_{3,4,6}$ asymmetries vanish in the large $z_2$ region,
 since the interference between the different helicity states of the $W$
 boson is very small.

On the other hand, the smaller $z_2$ becomes, the larger the gluon
 contribution becomes.
Due to this gluon contribution, the decay to the right-handed
 $W$ boson is allowed, even in the $m_b=0$ limit, which causes the
 deviation from the values in the two-body decay process.

\subsection{$T$-odd distributions}

Let us now turn to the $T$-odd asymmetries, the main subject of this
 article.
As mentioned above, the leading contribution to the $T$-odd effects in
 the top-quark decay (\ref{process}) comes from the interference
 between the tree diagrams and the absorptive part of the six one-loop
 diagrams in Fig.~\ref{feynman}.
The one-loop amplitudes are calculated in the $m_b=0$ limit, however the
 kinematical boundary as well as the cuts are given for $m_b=4$ GeV.
We set the QCD coupling constant as
 $\alpha_s=\alpha_s(k_{T {\rm min}}\!=\!20\,{\rm GeV})=0.15$.
The details of our calculation of the one-loop amplitudes are given in
 the appendices.

Figure~\ref{z2dis_todd} shows the asymmetry distributions as
 Fig.~\ref{z2dis_teven}, but for $A_{7,8,9}$ in Eq.~(\ref{asym}).
We found that the asymmetry $A_7$ is positive at a few percent level,
 and tends to be larger with increasing $z_1$ and decreasing $z_2$.
$A_{8}$ is also positive but less than 1\% in magnitude, and is large
 for the intermediate values of $z_1$ and $z_2$.
$A_{9}$ is the smallest in magnitude and is order permill.
It takes positive value for large $z_1$ and small $z_2$, but changes
 the sign by decreasing $z_1$ and increasing $z_2$.
The dips which appear in the figure are caused by the kinematical cuts
 given in Eqs.~(\ref{kt}) and (\ref{cosbg}).\\

\EPSFIGURE[t]{z2dis_todd,width=1.\textwidth,clip}
{The same as Fig.~\ref{z2dis_teven}, but for the $T$-odd asymmetries
 $A_{7,8,9}$ at the one-loop level.\label{z2dis_todd}}

In Fig.~\ref{A7dia}, we show the contribution to the $A_7$ asymmetry
 for $0.55<z_1<{z_1}_{\rm max}$ from the individual one-loop diagrams of
 Fig.~\ref{feynman} in the Feynman gauge.
The sum of the diagrams ($c$) and ($f$), which have the gluon
 three-point-vertex, gives negative contribution to $A_7$.
On the other hand, all the other diagrams give positive contribution to
 the asymmetry.
The diagrams ($a$) and ($d$) with $s$-channel $b$-quark exchange
 diagrams give the dominant contribution, which make the total asymmetry
 positive.
The diagrams ($b$) and ($e$), which contain the $u$-channel $b$-quark
 exchange in the final-state rescattering, are found to give small
 contribution.
On the other hand, the main contribution for $A_8$ comes from the
 diagrams ($c$)+($f$), while for $A_9$, the contributions from
 ($a$)+($d$) and ($c$)+($f$) are comparable in size.

\EPSFIGURE[l]{z2dis_A7,width=.325\textwidth,clip}
{(Left) The contribution to the $A_7$ asymmetry for
 $0.55\!<\!z_1\!<\!{z_1}_{\rm max}$ from the individual one-loop
 diagrams.
($a$)+($d$), ($b$)+($e$) and ($c$)+($f$) contributions in Feynman gauge
 are plotted in dashed, dotted and dotted-dashed line.
Total asymmetry is also plotted in solid line, as a reference.
\label{A7dia}}

\subsection{Up-down asymmetry for the LHC experiment}

In order to help finding an evidence of the $T$-odd asymmetries in
 experiments, we discuss a simple observable for the $T$-odd
 asymmetry.
We define the up-down asymmetry $A_{\rm UD}$ with respect to the top
 decay plane as
\begin{align}
 A_{\rm UD} \equiv \left[N(0<\phi<\pi)
 -N(\pi<\phi<2\pi)\right]/N_{\rm sum}. \label{aud}
\end{align}
It is defined as the asymmetry between the number of events having
 the charged lepton momentum with positive and negative $y$ component.
$A_{\rm UD}$ reflects the property of $A_{7}$, since
 $\sin\theta\sin\phi$ is positive for $0<\phi<\pi$
 while negative for $\pi<\phi<2\pi$.\\

We estimate $A_{\rm UD}$, and its statistical errors for 820,000
 top-quark signal events which is expected at the LHC
 one-year run with $L= 10$ fb$^{-1}$ after the event selection for the
 single lepton plus jets channel $pp\to t\bar t\to b\bar bWW\to b\bar
 b(\ell\nu)(jj)$~\cite{ATLAS}.
Taking into account the fraction%
\footnote{For the total decay width of the top quark, we use the
 calculation including the ${\cal O}(\alpha_s)$ QCD
 corrections~\cite{Czarnecki:1998qc}.}
 of $t\to bWg$ events that satisfy the kinematical cuts in
 Eqs.~(\ref{kt}) and (\ref{cosbg}), a sample of about 72,000 events for
 $t\to bWg$ followed by $W\to\ell\nu$ would be expected.
In Fig.~\ref{simu} (left), we display the distribution of the event
 sample in the $z_1$-$z_2$ plane. In order to see the $T$-odd
 asymmetries effectively, we divide the kinematical region into eight
 bins using the jet-energy ordering and the opening angle between the
 two jets in the top rest frame as
\begin{align}
   &{\rm (I)}&     &z_1>z_3&  &\quad\cos\theta_{bg}<-0.5, \quad
 & &{\rm (V)}&     &z_1<z_3&  &\quad\cos\theta_{bg}<-0.5, \nn\\
   &{\rm (II)}&    &z_1>z_3&  -0.5&<\cos\theta_{bg}< 0,   \quad
 & &{\rm (VI)}&    &z_1<z_3&  -0.5&<\cos\theta_{bg}< 0,   \nn\\
   &{\rm (III)}&   &z_1>z_3&     0&<\cos\theta_{bg}< 0.5, \quad
 & &{\rm (VII)}&   &z_1<z_3&     0&<\cos\theta_{bg}< 0.5, \nn\\
   &{\rm (IV)}&    &z_1>z_3&   0.5&<\cos\theta_{bg},      \quad
 & &{\rm (VIII)}&  &z_1<z_3&   0.5&<\cos\theta_{bg}. \label{div}
\end{align}
In the figure, the number of events in each bin are given in an unit of
 thousands.
As in Fig.~\ref{contour}, a large number of events is expected for the
 region where both $z_1$ and $z_2$ are large, namely (III) and (IV).

The top and middle plots in Fig.~\ref{simu} (right) show the up-down
 asymmetries with expected statistical error-bars for each of the eight
 bins, for the LHC one-year run.
The error is estimated from $\delta{A}=\sqrt{(1-A^2)/N_{\rm sum}}$ for
 each bin.
The magnitude of the asymmetry is larger for the (I)-(IV) bins than for
 the (V)-(VIII) bins, and increases with the opening angle
 $\theta_{bg}$, as is expected from the $z_1$ and $z_2$ dependences of
 $A_7$ in Fig.~\ref{z2dis_todd}.
The asymmetry reaches 3\% at the bin-(I) where, however, the
 event yield is not high.

In the bottom plot in Fig.~\ref{simu} (right),
 we also consider the case where the top-pair productions are identified
 without a $b$-jet-tagging.
In this case, instead of defining $y$-axis by the direction
 $\vec{q}\times\vec{p}_b$, we define the $y$-axis along
 $\vec{q}\times\vec{p}_{j_1}$, where $p_{j_1}$ is the momentum of the
 jet whose energy is large than the other in the top-quark rest frame.
This asymmetry corresponds to $A_{\rm UD}$ for $z_1>z_3$ (top) minus
 $A_{\rm UD}$ for $z_1<z_3$ (middle).
Because of the cancellation, the magnitude of the asymmetry decreases,
 but it remains finite even without $b$-jet identifications.

\FIGURE[t]{\epsfig{file=cont_event.eps,height=.295\textheight,clip}\quad
\epsfig{file=AUD.eps,height=.3\textheight,clip}
\caption{(Left) Estimation of the event yields for the LHC one-year run
 is shown in each bin defined in (\ref{div}).
 (Right) Up-down asymmetries $A_{\rm UD}$ defined in Eq.~(\ref{aud}) for
 the eight bins (top and middle) and $A_{\rm UD}$ for the case
 without $b$-tagging (bottom).
 $\cos\theta_{bg}$ is the opening angle between the two jets in the top
 rest frame.
 Error bars are estimated for the expected event yields
 shown in the left figure.\label{simu}}}
%

\section{Polarized top-quark decays}\label{sec:spin}

Although we have considered the decay of unpolarized top-quarks so far,
 the top-quarks produced singly by the electroweak interactions at
 hadron colliders or the top-quark pairs produced in $e^{+}e^{-}$
 colliders can be highly polarized.
Therefore, it may be useful to analyze the polarized top-quark decay.

In this section, we show that, when a top-quark is polarized, i) there
 exists another type of $T$-odd observable, the angular correlation
 between the top-spin direction and the top decay plane, and ii) the
 lepton angular distributions discussed in the previous section are
 modified\footnote{%
We thank the referee of this article for pointing out the existence of
 another $T$-odd observable in the polarized top-quark decay, and
 suggesting its relation to the normal polarization in the top-quark
 pair-production.}.\\

\EPSFIGURE[r]{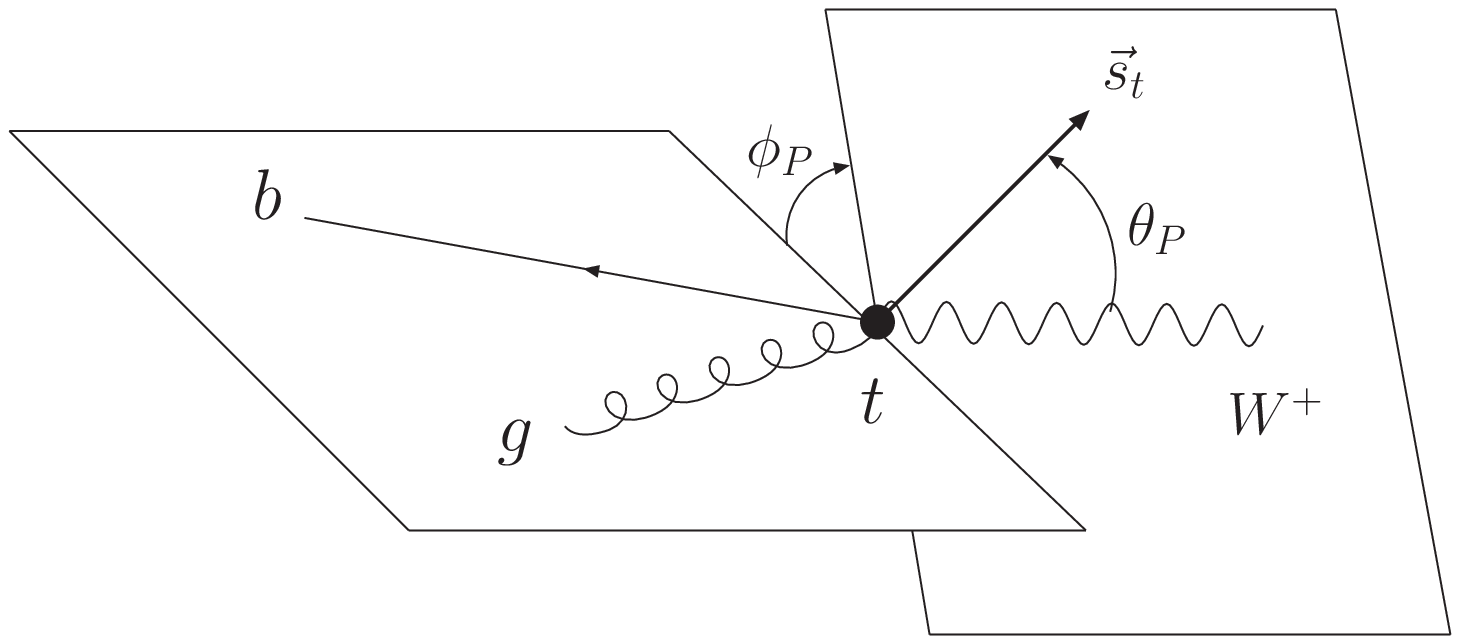,width=.5\textwidth,clip}
 {Schematic view of the coordinate system for the $t\to bW^{+}g$ decay,
 where the $W^+$ momentum direction in the top-quark rest-frame is
 chosen along the $z$-axis, with the top-quark's spin
 $\vec{s}_t$. \label{frame_spin}}

First, we discuss another type of $T$-odd observable in radiative decays
 of the polarized top-quarks, namely, the angular correlation between the
 top-quark spin and the decay plane.

We define the angles between the top-quark spin direction and the decay
 plane in the top-quark rest-frame as shown in Fig.~\ref{frame_spin}.
The $z$-axis is chosen along the $W$-momentum direction, and the
 $x$-axis is chosen along the $\vec{p}_b$ direction in the
 $(\vec{p}_b,\vec{p}_g)$ plane.
The polar and azimuthal angles, $\theta_{P}$ and $\phi_P$, respectively,
 define the direction of the top-quark spin $\vec{s}_t$.

The decay distribution is now characterized by the two angles as well as
$z_1$ and $z_2$:
\begin{align}
 \frac{d\Gamma}{dz_1dz_2d\cos{\theta_P}d\phi_P}
=&\frac{G_F\alpha_sm_t^3x^2|V_{tb}|^2C_F}{64\sqrt{2}\pi^3}\nn\\
&\times\Bigg[F_{P1}+F_{P2}\cos{\theta_P}+F_{P3}\sin{\theta_P}\cos{\phi_P}
+F_{P4}\sin{\theta_P}\sin{\phi_P}\Bigg].\label{spindist}
\end{align}
The structure functions $F_{P1-P4}(z_1,z_2)$ are obtained from the
 $t\to bWg$ matrix elements ${\cal M}_{\sigma_t}$, which are defined in
 Eq.~(\ref{amp}), but we now retain the top-quark helicity $\sigma_t$
 instead of the $W$-helicity ($\lambda$):
\begin{align}
 &F_{P1} = \frac{1}{2}{\textstyle \overline{\sum}}
\left(|{\cal M}_{+}|^2 + |{\cal M}_{-}|^2\right),
 &&F_{P3} = \frac{1}{2}{\textstyle \overline{\sum}}
\left({\cal M}^{*}_{+}{\cal M}_{-} +
{\cal M}^{*}_{-}{\cal M}_{+}\right),
\nonumber\\
 &F_{P2} = \frac{1}{2}{\textstyle \overline{\sum}}
\left(|{\cal M}_{+}|^2 - |{\cal M}_{-}|^2\right),
 &&F_{P4} = \frac{i}{2}{\textstyle \overline{\sum}}
\left({\cal M}^{*}_{+}{\cal M}_{-} -
{\cal M}^{*}_{-}{\cal M}_{+}\right).\label{fp}
\end{align}
The summation stands for the sum of the spins of all the particles
 but the top-quark, and the sum/average of colors.
The spin-independent term $F_{P1}$ is identical to $F_{1}$ in
 Eq.~(\ref{F1to9}), including the normalization factor.
$F_{P1}$ is $T$-even and $P$-even, while $F_{P2}$ and $F_{P3}$ are
 $T$-even and $P$-odd.
$F_{P4}$ is $T$-odd and $P$-even.
The leading-order contribution to the functions $F_{P1}$ to $F_{P3}$
 comes from the tree-level amplitudes.
On the other hand, the leading-order contribution to $F_{P4}$ comes from
 the interference between the tree amplitudes and the absorptive part of
 the one-loop amplitudes, just the same as $F_{7,8,9}$ in
 Eq.~(\ref{F1to9}).
Note that $F_{P4}$ is proportional to the expectation value of the
 triple product of the three vectors
 $\langle\vec{s}_t\cdot\vec{q}\times\vec{p}_b\rangle$, just like $F_{7}$
 is proportional to $\langle \vec{s}_W\cdot\vec{q}\times\vec{p}_b\rangle$.
The corresponding distribution for the anti-top-quark decay can be
 obtained by reversing the sign of $F_{P2}$ and $F_{P3}$ in
 Eq.~(\ref{spindist}), when the $CP$ is a good symmetry.\\

\EPSFIGURE[t]{z2dis_tpol,width=1.\textwidth,clip}
{The $z_2$ distributions of the angular correlation functions defined in
 Eq.~(\ref{ap}), $A_{P2,P3}$ at the tree level and $A_{P4}$ at the
 one-loop level, where the three $z_1$ regions and the kinematical cuts
 are the same as in Fig.~\ref{z2dis_teven}. \label{z2dis_tpol}}

We define the ratios of the correlation functions $F_{i}$ for
 $i=P2$-$P4$ to the spin-independent term $F_{P1}$ as
\begin{align}
 A_{i}(z_2) = \int dz_1 F_{i}(z_1,z_2)\bigg/\int dz_1 F_{P1} (z_1,z_2).
\label{ap}
\end{align}
Each correlation function corresponds to the expectation value of the
 component of the top-quark spin-vector as
\begin{align}
 \langle \vec s_t
 \rangle=\frac{1}{3}\left(A_{P3},\,A_{P4},\,A_{P2}\right).\label{spin}
\end{align}

In Fig.~\ref{z2dis_tpol}, the $z_2$ distributions of $A_{P2,P3}$ at the
 tree level and $A_{P4}$ at the one-loop level are shown, where the
 three $z_1$ regions and the kinematical cuts are the same as those in
 Fig.~\ref{z2dis_teven}.
The $T$-even $P$-odd asymmetries $A_{P2}$ and $A_{P3}$ are as large as a
 few times 10\% in magnitude, while the $T$-odd $P$-even asymmetry
 $A_{P4}$ is less than 1\%.
This means that the top-quark spin lies almost in the decay plane, or,
 the decay plane tends to contain the top-quark spin.
The $z_1$ dependence of $A_{P4}$ is similar to the $T$-odd lepton
 angular asymmetry $A_7$ in Fig.~\ref{z2dis_todd}.\\

\EPSFIGURE[r]{z2dis_A7_pol,width=.32\textwidth,clip}
{$T$-odd asymmetry $A_{7}$ for $0.55\!<\!z_1\!<\!{z_{1}}_{\rm max}$ in
 the polarized top-quark decays.
The direction of the polarization is parameterized from the $W$-momentum
 direction, $\theta_P=0^{\circ}$ (dashed) and $180^{\circ}$ (dotted).
As a reference, the unpolarized case is also
 plotted in solid line.\label{A7pol}}
Next, we consider the $T$-odd lepton angular asymmetry $A_7$ again, but
 in the decay of polarized top-quarks.
Since the degree of the normal polarization to the decay plane is quite
 small as shown in Fig.~\ref{z2dis_tpol},
for simplicity, the case that the top-quark spin lies in the decay plane,
 $\phi_P=0^{\circ}$, is considered.
In Fig.~\ref{A7pol}, we show the $A_{7}$ asymmetry for
 $0.55<z_1<{z_1}_{\rm max}$, where the spin direction of the top-quark
 is set at $\theta_{P}=0^{\circ}$ and $180^{\circ}$.
The asymmetry is enhanced when $\theta_P=0^{\circ}$, but
 reduced when $\theta_{P}=180^{\circ}$.
It follows from the fact that the decay amplitude to the
 right-handed $W$-boson is larger for $\theta_{P}=0^{\circ}$ than
 for $\theta_{P}=180^{\circ}$.\\

Finally, we briefly mention $T$-odd effects induced by the absorptive
 part of the top-pair production amplitudes, which produce the normal
 polarization with respect to the scattering plane.
The one-loop calculations have been done for $e^{+}e^{-}$
 and hadron colliders~\cite{Kuhn:1985ps,Kane:1991bg,Bernreuther:1992ef,%
Bernreuther:1995cx,Dharmaratna:1996xd,Brandenburg:1998xw}, however, the
 degree of polarization is estimated to be quite small.

We examine if the up-down asymmetry with respect to the decay plane of
 the top-quarks, studied in this paper, can contribute to the $T$-odd
 asymmetry about the scattering plane in the top-pair production process,
 when the production and decay processes are considered in total.
When the top-quark has normal polarization with respect to the
 scattering plane, because the charged lepton prefers to be emitted in
 the direction of the top-quark spin, the expectation value of the
 inner product of the top-quark spin direction and the lepton direction
 $\langle\vec{s}_t\cdot\vec{p}_{\ell}\rangle$ is positive.
On the other hand, considering the $T$-odd effects in the top-decay
 process, since the $A_{P4}$ asymmetry in Fig.~\ref{z2dis_tpol}
 is slightly positive, the expectation value of the triple product
 $\langle\vec{s}_t\cdot\vec{q}\times\vec{p}_b\rangle$ is slightly
 positive.
In addition, since $A_{7}$ in Fig.~\ref{z2dis_todd} is positive, the
 expectation value of
$\langle\vec{p}_\ell\cdot\vec{q}\times\vec{p}_b\rangle$ is also
 positive.
Therefore, the $T$-odd effect in the top decay process gives positive
 correction to $\langle\vec{s}_t\cdot\vec{p}_{\ell}\rangle$, i.e.\
 additive to the original asymmetry due to the $T$-odd polarization of
 the top-quark normal to the scattering plane.
However the size should be negligible, because the degree of the normal
 polarization and the $T$-odd correlation $A_{P4}$ are estimated to be
 very small.

Similarly, we find that the $T$-odd effect in the top-quark production
 process provides additive but negligible contribution to the $T$-odd
 asymmetry in the decay process with respect to the top decay plane.

\section{Summary}\label{sec:sum}

In this article, we studied the top quark decay into a bottom quark and
 a $W$ boson accompanied by one gluon emission, and calculated the
 absorptive part of the $t\to bWg$ decay amplitudes at the one-loop
 level.
We then estimated the leading-order contribution to the $T$-odd
 asymmetries of the lepton angular distribution in the $t\to bWg$ decay
 followed by leptonic decay of the $W$ boson.

For completeness, we also discussed the $T$-even asymmetries at the tree
 level ${\cal O}(\alpha_s)$, and found that the fraction to the
 right-handed $W$ boson increases in the small $W$-boson energy region.
As for the $T$-odd asymmetries, the largest asymmetry is predicted for
 $A_7$ at a few percent level, and the other asymmetries ($A_{8}$ and
 $A_{9}$) are found to be less than 1\%.

We proposed a simple observable $A_{\rm UD}$, the up-down asymmetry with
 respect to the top decay plane, which is proportional to $A_7$.
The $A_{\rm UD}$ asymmetry is predicted to be at a few percent level,
 which may be confirmed at the LHC with 10 fb$^{-1}$.

Before closing let us mention that, for the polarized top-quark decays,
 there exists another $T$-odd observable, the angular correlation
 between the top-quark spin direction and the top decay plane.
However, the size of the $T$-odd correlation is less than 1\%, which may
 be difficult to measure at future colliders.

\acknowledgments

K.H. wishes to thank E.~Asakawa and M.~Tanaka for discussions in the
 early stage of the investigation.
K.M.\ and H.Y.\ thank the KEK theory group for the warm hospitality.
We thank B.~J\"ager, Y.~Kurihara and E.~Senaha for discussions.
This work is supported in part by the core university program of JSPS,
 and in part by the Grant-in-Aid for Scientific Research (Nos. 17540281,
 18340060) of MEXT, Japan.
The work of H.Y.\ is supported in part by a Research Fellowship
 of the Japan Society for the Promotion of Science.

\appendix

\section{$t\to bW^+g$ decay amplitudes}\label{app:amp}

In this appendix, we outline our calculation of the amplitude
 for the $t\to bWg$ process.
Note that we present the formalism in the $m_b=0$ limit, because we
 performed the one-loop calculation only in this limit.
The extension to the massive $b$-quark case will be given only for the
 tree-level calculation.\\

First, we expand the tensor $T^{\ma}$ in Eq.~(\ref{amp}) as
\begin{align}
T^{\ma}=\sum_{i}a_{i}\,T^{\ma}_{i}\label{sum}
\end{align}
 with the 20 basis tensors;
\begin{align}
&\TL{1}  = g^{\ma}q\dsl P_{-}/m_t^2,&
&\TL{6}  = \gamma^{\mu}p_b^{\alpha}P_{-}/m_t^2,&
&\TR{1}  = g^{\ma}P_{+}/m_t,&
&\TR{6}  = \gamma^{\mu}p_b^{\alpha}q\dsl P_{+}/m_t^3,\nn\\
&\TL{2}  = \gamma^{\mu}\gamma^{\alpha}q\dsl P_{-}/m_t^2,&
&\TL{7}  = p_t^{\mu}p_t^{\alpha}q\dsl P_{-}/m_t^4,&
&\TR{2}  = \gamma^{\mu}\gamma^{\alpha}P_{+}/m_t,&
&\TR{7}  = p_t^{\mu}p_t^{\alpha}P_{+}/m_t^3,\nn\\
&\TL{3}  = p_t^{\mu}\gamma^{\alpha}P_{-}/m_t^2,&
&\TL{8}  = p_t^{\mu}p_b^{\alpha}q\dsl P_{-}/m_t^4,&
&\TR{3}  = p_t^{\mu}\gamma^{\alpha}q\dsl P_{+}/m_t^3,&
&\TR{8}  = p_t^{\mu}p_b^{\alpha}P_{+}/m_t^3,\nn\\
&\TL{4}  = p_b^{\mu}\gamma^{\alpha}P_{-}/m_t^2,&
&\TL{9}  = p_b^{\mu}p_t^{\alpha}q\dsl P_{-}/m_t^4,&
&\TR{4}  = p_b^{\mu}\gamma^{\alpha}q\dsl P_{+}/m_t^3,&
&\TR{9}  = p_b^{\mu}p_t^{\alpha}P_{+}/m_t^3,\nn\\
&\TL{5}  = \gamma^{\mu}p_t^{\alpha}P_{-}/m_t^2,&
&\TL{10} = p_b^{\mu}p_b^{\alpha}q\dsl P_{-}/m_t^4,&
&\TR{5}  = \gamma^{\mu}p_t^{\alpha}q\dsl P_{+}/m_t^3,&
&\TR{10} = p_b^{\mu}p_b^{\alpha}P_{+}/m_t^4&
\label{list}
\end{align}
 where the chiral-projection operators are
 $P_{\pm}=\frac{1}{2}(1\pm\gamma_5)$.
 The summation runs for $i=\{L1$-$L10$,$R1$-$R10\}$.
The coefficients $a_{i}$ are calculated perturbatively,
\begin{align}
a_{i} = b_{i} + i\alpha_{s} c_{i} + \cdots,\label{ai}
\end{align}
 where $b_{i}$ is the tree-level contribution, and $c_{i}$ is the
 one-loop contribution to the absorptive part.

The 20 coefficients satisfy the following sum rules, because of the
gauge invariance of QCD (${p_g}_{\alpha} T^{\ma}=0$),
\begin{align}
&2(1-y_2)a_{L2}+z_3a_{L5}+y_2a_{L6}+2a_{R2}=0,\nn\\
&2a_{L2}+2a_{R2}-z_3a_{R5}-y_2a_{R6}=0,\nn\\
&2a_{L1}-2a_{L3}+z_3a_{L7}+y_2a_{L8}-2a_{R3}=0,\nn\\
&2a_{L3}+2a_{R1}+2(1-y_2)a_{R3}+z_3a_{R7}+y_2a_{R8}=0,\nn\\
&2a_{L1}+4a_{L2}+2a_{L4}-z_3a_{L9}-y_2a_{L10}+2a_{R4}=0,\nn\\
&2a_{L4}-2a_{R1}-4a_{R2}+2(1-y_2)a_{R4}+z_3a_{R9}+y_2a_{R10}=0,
\label{gauge}
\end{align}
 where we defined $y_2=1\!-\!z_2\!+\!x^2$.
In appendices \ref{sub:tree}, \ref{sub:one} and \ref{app:loop}, we
 present the following 14 coefficients;
$i=L1$-$L4$,$L6$,$L8$,$L10$,$R1$-$R4$,$R6$,$R8$,$R10$.
The remaining 6 coefficients; $i=L5$,$L7$,$L9$,$R5$,$R7$,$R9$
 are then obtained from the above identities, Eq.~(\ref{gauge})%
\footnote{
To verify our results, we have calculated all the 20 coefficients
independently and checked that these satisfy the Eqs.~(\ref{gauge}).
}.

Counting the number of the physical amplitudes, only the twelve among
 the 14 coefficients are independent~\cite{perrottet,Hagiwara:1982cq}.
Using the Dirac matrix identity~\cite{perrottet}, the terms with
 $T^{\ma}_{L10}$ and $T^{\ma}_{R10}$ can be removed by the following
 replacements,
\begin{align}
&a_{L1} \to a_{L1}+\frac{1}{2}(y_3-z_1z_2)a_{L10}
 -\frac{1}{2}z_1a_{R10},\nn\\
&a_{L2} \to a_{L2}-\frac{1}{2}(y_3-z_1z_2)a_{L10}
 +\frac{1}{2}z_1a_{R10},\nn\\
&a_{L3} \to a_{L3}-\frac{1}{2}z_2y_3a_{L10}
 -\frac{1}{2}y_3a_{R10},\nn\\
&a_{L4} \to a_{L4}+\frac{1}{2}(z_2^2-2x^2)a_{L10}
 +\frac{1}{2}z_2a_{R10},\nn\\
&a_{L6} \to a_{L6}+\frac{1}{2}\{(z_1-z_2)z_2+2x^2\}a_{L10}
 +\frac{1}{2}(z_1-z_2)a_{R10},\nn\\
&a_{L8} \to a_{L8}+z_2a_{L10}+a_{R10},\nn\\
&a_{R1} \to a_{R1}+\frac{1}{2}z_1x^2a_{L10}+\frac{1}{2}y_3a_{R10},\nn\\
&a_{R2} \to a_{R2}-\frac{1}{2}z_1x^2a_{L10}-\frac{1}{2}y_3a_{R10},\nn\\
&a_{R3} \to a_{R3}+\frac{1}{2}y_3a_{L10},\nn\\
&a_{R4} \to a_{R4}-\frac{1}{2}z_2a_{L10}-a_{R10},\nn\\
&a_{R6} \to a_{R6}-\frac{1}{2}(z_1-z_2)a_{L10}+a_{R10},\nn\\
&a_{R8} \to a_{R8}-x^2 a_{L10},
\end{align}
with $y_3=1-z_3-x^2$.

\subsection{Tree-level results}\label{sub:tree}

At the tree level, the amplitude has the contributions from two Feynman
 diagrams (Fig.~\ref{feynman}),
\begin{align}
T_{\rm tree}^{\ma}=
\gamma^{\alpha}\frac{1}{p\dsl_t-q\dsl+i\epsilon}
\gamma^{\mu}P_{-}+\gamma^{\mu}P_{-}
\frac{1}{p\dsl_t-p\dsl_g-m_t+i\epsilon}
\gamma^{\alpha}.   \label{treeamp}
\end{align}
The decomposition in terms of $T^{\ma}_{i}$ in (\ref{list}) gives
\begin{align}
 &b_{L1}=b_{L3}
=-b_{R1}=2x_b,
\aki b_{L4}
=-b_{L6}=2x_t, \aki -b_{L2}=b_{R2}=x_t+x_b,
\end{align}
 where $x_t\equiv{m_t^2}/{(-2p_t\cdot p_g)}$ and
 $x_b\equiv{m_t^2}/{2p_b\cdot p_g}$.\\

For the massive $b$-quark case, two more components,
\begin{align}
 T^{\ma}_{M1}=g^{\ma}P_{-}/m_t,\quad
 T^{\ma}_{M2}=\gamma^{\mu}\gamma^{\alpha}P_{-}/m_t,
\end{align}
 with the coefficients $b_{M1}=2yx_b$, $b_{M2}=-y(x_t+x_b)$ and
 $y=m_b/m_t$, must be added to Eq.~(\ref{sum}).

\subsection{One-loop results}\label{sub:one}

At the one-loop level, the absorptive part emerges from the six
 diagrams for the $t\to bWg$ decay, shown in Fig.~\ref{feynman}.
We write the one-loop coefficients in Eq.~(\ref{ai}) as the sum of these
 diagrammatic contributions,
\begin{align}
 c_{i}= c_{i}^{(a)}+c_{i}^{(b)}+c_{i}^{(c)}
       +c_{i}^{(d)}+c_{i}^{(e)}+c_{i}^{(f)}. \label{ci}
\end{align}
The analytic expressions of the coefficients are obtained for each
 diagram by performing the standard Feynman integrals.
Our expression contains functions with a one-parameter integral,
 which can easily be evaluated.

In the next appendix, we also show the results of $c_i$ in the loop
 scalar function method as an alternative expression.
We checked that the numerical results of the two calculations agree
 completely.\\

With the color factor $C_F=4/3$, $C_A=3$ and $C_1=C_F-C_A/2=-1/6$,
 the one-loop coefficients for each diagram are found as below;\\

\noindent
$\bullet$ diagram-($a$)
\begin{align}
 -c^{(a)}_{L1} =2c^{(a)}_{L2}=-c^{(a)}_{L3}
 =c^{(a)}_{R1}=-2c^{(a)}_{R2}=\frac{C_F}{2} x_b.
\end{align}

\noindent
$\bullet$ diagram-($b$)
\begin{align}
-2c^{(b)}_{L1}=4c^{(b)}_{L2}=-2c^{(b)}_{L3}
=c^{(b)}_{L6}=2c^{(b)}_{R1}=-4c^{(b)}_{R2}=C_1 x_b.
\end{align}

\noindent
$\bullet$ diagram-($c$)
\begin{align}
-c^{(c)}_{L1}=2c^{(c)}_{L2}=-c^{(c)}_{L3}
=2c^{(c)}_{L6}=c^{(c)}_{R1}=-2c^{(c)}_{R2}=\frac{C_A}{2}
 x_b\left(\ln{\epsilon^2}+\ln{x_b}\right),
\end{align}
 where $\epsilon = m_g/m_t$.
The gluon mass $m_g$ is introduced to regulate the IR singularity.
We keep $\epsilon$ only in the singular parts and take the
 $\epsilon\to 0$ limit elsewhere.

\noindent
$\bullet$ diagram-($d$)
\begin{align}
&c^{(d)}_{L1}=-2c^{(d)}_{L2}=-C_F\left[x_b(2-z_2)I_{10}-(3-z_2)I_{21}
 +\frac{2-y_2}{2}I_{22}\right],\nn\\
&c^{(d)}_{L3}=-C_F\left[x_b(1-x^2)I_{10}-(1-x^2)I_{21}
-\frac{y_2}{2}I_{22}-x^2y_2\left(I_{33}-I_{34}\right)\right],\nn\\
&c^{(d)}_{R1}=-2c^{(d)}_{R2}=C_F\left[x_b(1-x^2)I_{10}-(2-x^2)I_{21}
+\frac{2-y_2}{2}I_{22}\right],\nn\\
&c^{(d)}_{R3}=C_F\left[I_{21}-I_{22}-y_2\left(I_{32}-I_{33}\right)\right].
\end{align}
Here, $I_{mn}$ is defined by the integral
\begin{align}
 I_{mn}=\int_0^1\frac{t^n\,dt}{[1-z_2 t+ x^2 t^2]^{m}}.\label{mn}
\end{align}

\noindent
$\bullet$ diagram-($e$)
\begin{align}
&c^{(e)}_{L1}=C_1\left[x_b\ln(z_1^2 x_b)-(1+z_1)J_{110}
+(z_2-x^2)J_{111}+\frac{y_2^2}{2}J_{213}
-z_1 L_2\right],\nn\\
&c^{(e)}_{L2}=-\frac{1}{2}c^{(e)}_{L1}+\frac{C_1}{2}y_2\left[
z_1(J_{121}-J_{122})+J_{211}-2J_{212}
+\frac{2-y_2}{2}J_{213}\right],\nn\\
&c^{(e)}_{L3}=C_1\left[x_b\ln(z_1^2 x_b)-I_{10}+y_2I_{21}
-(1+z_1)J_{110}+(z_2-x^2)J_{111}+y_2^2J_{121}\right.\nn\\
&\left.\hspace{50pt}
+y_2J_{211}-\frac{y_2(2-y_2)}{2}J_{212}-z_1 L_2\right],\nn\\
&c^{(e)}_{L4}=C_1\left[J_{110}-(z_2-x^2)J_{111}
-y_2(1-z_2)J_{121}-\frac{y_2(y_2+2x^2)}{2}J_{122}\right],\nn\\
&c^{(e)}_{L6}=-C_1\left[2x_b+J_{110}-(1+x^2)J_{111}
-y_2(2+z_1-z_2)J_{121}+\frac{y_2(2-y_2)}{2}J_{122}
\right.\nn\\&\left.\hspace{50pt}
+\frac{y_2(y_2+2x^2)}{2}J_{213}
+\frac{z_1(z_1+y_2)}{z_1-y_2}L_2
-\frac{2z_1^2y_2}{z_1-y_2}L_3\right],\nn\\
&c^{(e)}_{L8}=-C_1\left[4J_{111}-2y_2\left(J_{121}+J_{122}
+2J_{212}+J_{213}\right)+y_2^2\left(J_{223}+2J_{314}\right)\right],\nn\\
&c^{(e)}_{L10}=C_1y_2\left[2J_{122}
-y_2\left(2J_{133}+J_{224}\right)\right],\nn\\
&c^{(e)}_{R1}=-c^{(e)}_{L1}+C_1\left[
I_{10}-y_2J_{110}
-\frac{y_2^2}{2}\left(J_{212}-J_{213}\right)\right],\nn\\
&c^{(e)}_{R2}=-\frac{1}{2}c^{(e)}_{R1}+\frac{C_1}{2}y_2\left[
-2J_{111}+y_2J_{121}+(2-z_2)J_{211}
-\frac{1+z_2-3x^2}{2}J_{212}\right],\nn\\
&c^{(e)}_{R3}=-C_1y_2\left[J_{211}-J_{212}\right],
\hspace{20pt}c^{(e)}_{R4}=-C_1\left[J_{110}-J_{111}
+y_2\left(J_{121}-J_{122}\right)\right], \nn\\
&c^{(e)}_{R6}=C_1\left[J_{110}-2J_{111}
+y_2\left(J_{121}+J_{212}\right)\right],  \nn\\
&c^{(e)}_{R8}=C_1\left[2J_{110}-2y_2\left(J_{121}+2J_{211}\right)
+y_2^2\left(J_{222}+2J_{313}\right)\right], \nn\\
&c^{(e)}_{R10}=-C_1y_2\left[4J_{121}-2J_{122}-y_2\left(
2J_{132}+J_{223}\right)\right],
\label{ce}
\end{align}
 where $J_{mn\ell}$ and $L_n$ are defined as
\begin{align}
&J_{mn\ell}=\int_0^1 \frac{t^\ell dt}
{[1-z_2 t+ x^2 t^2]^{m}[y_2t +z_1(1-t)]^{n}},\label{mnl}\\[2mm]
 &L_n=\int_0^1\frac{dt}{[y_2t+z_1(1-t)]^n}
\ln\left(\frac{y_2 t^2}{1-z_2 t+x^2 t^2}\right).\label{ln}
\end{align}

\noindent
$\bullet$ diagram-($f$)
\begin{align}
&c^{(f)}_{L1}=\frac{C_A}{2}\left[
x_b \ln{(z_3^2 x_b)}-x_b -(1+z_3)J'_{110}+\frac{1+x^2}{2}J'_{111}
+\frac{y_2(2+z_3)}{2}J'_{121}-\frac{z_2y_2}{2}J'_{122}
\right.\nn\\&\left.\hspace{50pt}
+y_2J'_{212}-\frac{y_2(1+x^2)}{2}J'_{213}
-\frac{z_3(y_2-3z_3)}{2(y_2-z_3)}L'_2
-\frac{y_2z_3^2}{y_2-z_3}L'_3
\right],\nn\\
&c^{(f)}_{L2}=\frac{C_A}{4}\left[
x_t\left(\ln{\epsilon^2}+\ln{x_b}\right)-x_b\ln{(z_3^2 x_b)}
+x_t+x_b+(1+z_3)J'_{110}-\frac{2-z_2+2x^2}{2}J'_{111}
\right.\nn\\&\left.\hspace{50pt}
-\frac{y_2(z_2+z_3)}{2}J'_{121}+x^2y_2J'_{122}
-\frac{y_2}{2}J'_{211}+\frac{x^2y_2}{2}J'_{213}
+\frac{3}{2}(y_2+z_3)L'_2-y_2z_3L'_3
\right],\nn\\
&c^{(f)}_{L3}=\frac{C_A}{2}\left[
x_b\ln{(z_3^2 x_b)}-\!x_b-\!2z_3J'_{110}\!
-\frac{z_2\!-\!2z_3\!-\!2x^2}{2}J'_{111}\!-y_2J'_{121}\!
+\frac{y_2(3z_2\!+\!z_3\!-\!4x^2)}{2}J'_{122}
\right.\nn\\&\left.\hspace{50pt}
-y_2(1-z_3)J'_{211}+\frac{y_2(3-2z_3-x^2)}{2}J'_{212}
-\frac{z_3(y_2-3z_3)}{2(y_2-z_3)}L'_2
-\frac{y_2z_3^2}{y_2-z_3}L'_3
\right],\nn\\
&c^{(f)}_{L4}=-\frac{C_A}{2}\left[
x_t\left(\ln{\epsilon^2}+\ln{x_b}\right)+x_t+J'_{110}-(z_2-x^2)J'_{111}
-\frac{y_2(2+z_2)}{2}J'_{121}
\right.\nn\\&\left.\hspace{50pt}
+\frac{y_2(1+2z_2-x^2)}{2}J'_{122}
+\frac{y_2(3y_2-z_3)}{2(y_2-z_3)}L'_2
-\frac{y_2^2z_3}{y_2-z_3}L'_3
\right],\nn\\
&c^{(f)}_{L6}=\frac{C_A}{2}\left[
(x_t-\frac{x_b}{2})\left(\ln{\epsilon^2}+\ln x_b\right)
+x_t+x_b+J'_{110}-\frac{1+z_2+x^2}{2}J'_{111}
-\frac{y_2(z_2+z_3)}{2}J'_{121}
\right.\nn\\&\left.\hspace{35pt}
+\frac{y_2(2z_2-z_3)}{2}J'_{122}
-\frac{y_2(2-z_2)}{2}J'_{212}+\frac{y_2(1+x^2)}{2}J'_{213}
+\frac{3y_2+2z_3}{2}L'_2
-y_2z_3L'_3
\right],\nn\\
&c^{(f)}_{L8}=\frac{C_A}{2}\left[
4J'_{111}-2y_2\left(3J'_{122}+2J'_{212}+J'_{213}\right)
+y_2^2\left(2J'_{133}+J'_{223}+J'_{224}+2J'_{314}\right)
\right],\nn\\
&c^{(f)}_{L10}=\frac{C_A}{2}y_2\left[
2J'_{122}-y_2\left(2J'_{133}+J'_{224}\right)
\right],\nn\\
&c^{(f)}_{R1}=-c^{(f)}_{L1}+\frac{C_A}{2}\left[
I_{10}-\frac{y_2}{2}\Big\{z_3(J'_{121}-J'_{122})
+J'_{211}-(2+y_2)J'_{212}+J'_{213}\Big\}
\right],\nn\\
&c^{(f)}_{R2}=-c^{(f)}_{L2}+\frac{C_A}{4}\left[
I_{10}-\frac{y_2+4z_3}{2}J'_{110}
-(y_2-2z_3)J'_{111}-\frac{y_2^2}{2}J'_{121}\right.\nn\\
&\left.\hspace{100pt}-\frac{y_2(1-z_2)}{2}J'_{211}
-\frac{x^2y_2}{2}\left(2J'_{212}-J'_{213}\right)
\right],\nn\\
&c^{(f)}_{R3}=-\frac{C_A}{2}\left[
J'_{110}-J'_{111}-2y_2\left(
J'_{121}-J'_{122}+J'_{211}-J'_{212}\right)
\right],\nn\\
&c^{(f)}_{R4}=\frac{C_A}{2}\left[J'_{110}-J'_{111}
-2y_2\left(J'_{121}-J'_{122}\right)\right],\nn\\
&c^{(f)}_{R6}=-\frac{C_A}{2}\left[
J'_{110}-2J'_{111}+y_2\left(J'_{122}+J'_{212}\right)
\right], \nn\\
&c^{(f)}_{R8}=-C_A\left[
J'_{110}-2y_2\left(J'_{121}+J'_{211}\right)
+y_2^2\left(J'_{133}+J'_{223}+J'_{313}\right)
\right],\nn\\
&c^{(f)}_{R10}=-\frac{C_A}{2}y_2\left[
4J'_{121}-2J'_{122}-y_2\left(2J'_{133}+J'_{223}\right)
\right],
\end{align}
 where $J'_{mn\ell}$ and $L'_{n}$ are given by
 replacing $z_1$ to $z_3$ in $J_{mn\ell}$ and $L_{n}$ in Eq.~(\ref{mnl})
 and Eq.~(\ref{ln}), respectively. \\

We note that the sum of the IR singular terms from the diagrams ($c$)
 and ($f$) is exactly proportional to the tree-level amplitude,
 therefore they do not contribute to the $T$-odd distribution.

\section{Loop scalar functions}\label{app:loop}

As a check of our calculation, we calculate the one-loop coefficients
 in terms of the loop scalar functions, the Passarino and Veltman's $B$,
 $C$, $D$ functions~\cite{Passarino:1978jh}.

For each diagram, we assign the masses and the momenta of the scalar
 function, following the FF notation~\cite{vanOldenborgh:1990yc}, and
 take only the imaginary part of the functions.
In this assignment we explicitly present the $b$-quark and gluon mass,
 $m_{b,g}$, for clarity, even though we take the massless limit in our
 analysis.\\

\noindent
$\bullet$ {diagram-($a$)}

Defining
 $B_{i}=\im\, B_i(m_g^2,m_b^2;\, p_{bg}^2)$ for $i$=0,1 with
 $p_{bg}^2=(p_b+p_g)^2$, the coefficients are expressed as
\begin{align}
 -c^{(a)}_{L1}=2c^{(a)}_{L2}=-c^{(a)}_{L3}=c^{(a)}_{R1}=-2c^{(a)}_{R2}
 =C_Fx_b\big[B_0+B_1\big]/\pi.
\end{align}

\noindent
$\bullet$ {diagram-($b$)}

Defining
 $C_{i}^{(b)}=\im\, C_i(m_g^2,m_b^2,m_b^2;\, p_{bg}^2,p_g^2,p_b^2)$
 for $i$=0,11,12,21-24, the coefficients are expressed as
\begin{align}
 -c^{(b)}_{L1}&=2c^{(b)}_{L2}=-c^{(b)}_{L3}=c^{(b)}_{R1}=-2c^{(b)}_{R2}
  \nn\\
 &=C_1 \big[-C_0-2C_{11}+C_{12}-C_{21}+C_{23}-2C_{24}/p_{bg}^2\big]\,
   m_t^2/\pi, \nn\\
 c^{(b)}_{L6}&=C_1 \big[-C_0-2C_{11}+C_{12}-C_{21}+C_{23}\big]\,
   m_t^2/\pi.
\end{align}

\noindent
$\bullet$ {diagram-($c$)}

Defining
 $C_{i}^{(c)}=\im\, C_i(m_g^2,m_b^2,m_g^2;\, p_{bg}^2,p_b^2,p_g^2)$
 for $i$=0,11,12,21-24, the coefficients are expressed as
\begin{align}
 -c^{(c)}_{L1}&=2c^{(c)}_{L2}=-c^{(c)}_{L3}=c^{(c)}_{R1}=-2c^{(c)}_{R2}
  \nn\\
 &=C_A \big[C_0-C_{11}-2C_{21}+2C_{23}-12C_{24}/p_{bg}^2\big]\,
   m_t^2/4\pi, \nn\\
 c^{(c)}_{L6}&=C_A
   \big[2C_0+4C_{11}-3C_{12}+2C_{21}-2C_{23}\big]\,m_t^2/4\pi.
\end{align}

\noindent
$\bullet$ {diagram-($d$)}

Defining
 $C_{i}^{(d)}=\im\, C_i(m_g^2,m_t^2,m_b^2;\, p_t^2,q^2,p_{bg}^2)$
 for $i$=0,11,12,21-24, the coefficients are expressed as
\begin{align}
 &c^{(d)}_{L1}=-2c^{(d)}_{L2}
 =-C_Fx_b\big[-(2-z_2)C_0-(3-z_2)C_{11}+(z_2-x^2)C_{12}
               -C_{21}-x^2C_{22} +z_2C_{23} \nn\\
 &\hspace*{4cm} -2C_{24}/m_t^2\big]\,m_t^2/\pi, \nn\\
 &c^{(d)}_{L3}
 =-C_Fx_b\big[-(1-x^2)(C_0+C_{11})-x^2(C_{22}-C_{23})-2C_{24}/m_t^2
          \big]\,m_t^2/\pi, \nn\\
 &c^{(d)}_{R1}=-2c^{(d)}_{R2}
 =-C_Fx_b\big[(1-x^2)C_0+(2-x^2)C_{11}-(z_2-x^2)C_{12}
               +C_{21}+x^2C_{22} -z_2C_{23} \nn\\
 &\hspace*{4cm} +2C_{24}/m_t^2\big]\,m_t^2/\pi, \nn\\
 &c^{(d)}_{R3}=-C_Fx_b\big[-C_{11}+C_{12}-C_{21}+C_{23}\big]\,m_t^2/\pi.
\end{align}

\noindent
$\bullet$ {diagram-($e$)}

Defining $D_{i}^{(e)}=\im\, D_i(m_g^2,m_t^2,m_b^2,m_b^2;\,
 p_t^2,q^2,p_g^2,p_b^2,p_{bg}^2,(q+p_g)^2)$ for
 $i$=0,11-13,21-27,31-313, the coefficients are expressed as
\begin{align}
 c^{(e)}_{L1}
 =&C_1\big[z_1D_0+(1+z_1)D_{11}-D_{12}+2(D_{27}+D_{312}-D_{313})/m_t^2
      \big]\,m_t^4/\pi, \nn\\
 c^{(e)}_{L2}
 =&C_1\big[-z_1D_0-(1+z_1)D_{11}+D_{12}
            +D_{21}+(z_2-x^2)D_{22}-(1-z_2+x^2)D_{23}-2D_{24} \nn\\
    &\quad +z_1D_{25}-(z_2-z_3-2x^2)D_{26}-2D_{27}/m_t^2
            -x^2D_{32}-D_{34}+D_{35}+z_2D_{36} \nn\\
    &\quad -z_3D_{37}-(1-z_1-2x^2)D_{38}+(1-z_1-x^2)D_{39}
           -(z_2-z_3)D_{310} \nn\\
    &\quad -6(D_{312}-D_{313})/m_t^2\big]\,m_t^4/2\pi, \nn\\
 c^{(e)}_{L3}
 =&C_1\big[ z_1D_0+(1+2z_1)D_{11}+(1-z_1-2z_2+2x^2)D_{12}
           -(1-z_2+x^2)(2D_{13}-D_{23}) \nn\\
    &\quad +(2+z_1)D_{21}+x^2D_{22}
          +(1-z_1-3z_2+2x^2)D_{24}-(4-2z_2+x^2)D_{25} \nn\\
    &\quad +(2z_2-3x^2)D_{26}+D_{31}-z_2D_{34}
           -(1+z_3)D_{35}+x^2(D_{36}-D_{38})+z_3D_{37} \nn\\
    &\quad -(1-z_1-x^2)D_{39}+(1-z_1+z_2-x^2)D_{310}
           +4(D_{27}+D_{311}-D_{313})/m_t^2\big]\,m_t^4/\pi, \nn\\
 c^{(e)}_{L4}
 =&C_1\big[-D_{11}+(z_2-x^2)D_{12}+(1-z_2+x^2)D_{13}
           -D_{21}-x^2D_{22}+z_2D_{24}+D_{25} \nn\\
    &\quad -(z_2-x^2)D_{26}-2(D_{27}+D_{313})/m_t^2\big]
          \,m_t^4/\pi, \nn\\
 c^{(e)}_{L6}
 =&C_1\big[ D_{11}+(1-2z_2+x^2)D_{12}+(z_2-z_3-x^2)D_{13}
           +2D_{21}+x^2D_{22} \nn\\
    &\quad -(1-z_2+x^2)D_{23} -2z_2D_{24}+(z_1-z_3)D_{25}
           +(1-z_1)D_{26}+D_{35}-z_3D_{37} \nn\\
    &\quad +x^2D_{38}+(1-z_1-x^2)D_{39}-z_2D_{310}
           +2(D_{27}-D_{312}+3D_{313})/m_t^2\big]\,m_t^4/\pi, \nn\\
  c^{(e)}_{L8}
 =&C_1\big[ 2(D_{12}-D_{13}+D_{24})+D_{22}+D_{23}-D_{25}-3D_{26}
           +D_{36}-D_{38}+D_{39}-D_{310}\big] \nn\\
    &\quad \times 2m_t^4/\pi, \nn\\
 c^{(e)}_{L10}
 =&C_1\big[-D_{23}+D_{26}+D_{38}-D_{39}\big]\,2m_t^4/\pi, \nn\\
 c^{(e)}_{R1}
 =&C_1\big[-z_1D_0-(2z_1+z_2-x^2)D_{11}-(1-z_1-2z_2+2x^2)D_{12}
           +(1-z_2+x^2)D_{13} \nn\\
    &\quad -D_{21}-x^2D_{22}+z_2D_{24}+z_3D_{25}-(1-z_1-x^2)D_{26} \nn\\
    &\quad -2(2D_{27}+D_{311}-D_{313})/m_t^2\big]\,m_t^4/\pi, \nn\\
 c^{(e)}_{R2}
 =&C_1\big[ z_1D_0+(2z_1+z_2-x^2)D_{11}+(1-z_1-2z_2+2x^2)D_{12}
           +(2-z_2)D_{21} \nn\\
    &\quad -(1-z_2+x^2)(D_{13}-D_{23})
           -(z_2-2x^2)(D_{24}-D_{26})-(3-2z_2+x^2)D_{25}+D_{31} \nn\\
    &\quad -z_2D_{34}-(1+z_3)D_{35}
           +x^2(D_{36}-D_{38})+z_3D_{37}-(1-z_1-x^2)D_{39} \nn\\
    &\quad +(1-z_1+z_2-x^2)D_{310}
           +2(2D_{27}+3D_{311}-3D_{313})/m_t^2\big]\,m_t^4/2\pi, \nn\\
 c^{(e)}_{R3}
 =&C_1\big[-D_{21}+D_{24}+D_{25}-D_{26}\big]\,m_t^4/\pi, \nn\\
 c^{(e)}_{R4}
 =&C_1\big[D_{11}-D_{12}-D_{25}+D_{26}\big]\,m_t^4/\pi, \nn\\
 c^{(e)}_{R6}
 =&C_1\big[-D_{11}+2D_{12}-D_{13}+D_{24}-D_{26}\big]\,m_t^4/\pi, \nn\\
 c^{(e)}_{R8}
 =&C_1\big[-D_{11}+D_{13}-2D_{21}-D_{23}+3D_{25}
           -D_{34}-D_{39}+2D_{310}\big]\,2m_t^4/\pi, \nn\\
 c^{(e)}_{R10}
 =&C_1\big[ D_{23}-2D_{25}+D_{26}+D_{39}-D_{310}\big]\,2m_t^4/\pi.
\end{align}

\noindent
$\bullet$ {diagram-($f$)}

Defining $D_{i}^{(f)}=\im\, D_i(m_g^2,m_t^2,m_b^2,m_g^2;\,
 p_t^2,q^2,p_b^2,p_g^2,p_{bg}^2,(q+p_b)^2)$ for
 $i$=0,11-13,21-27,31-313, the coefficients are expressed as
\begin{align}
 c^{(f)}_{L1}
 =&-C_A\big[-2z_3D_0-2(1+z_3)D_{11}-(2-3z_2+2x^2)D_{12}
            +(2+z_1-2z_2+2x^2)D_{13} \nn\\
     &\quad -3D_{21}+(z_2-3x^2)D_{22}-(2-3z_2)D_{24}+3z_1D_{25}
            +(3-2z_1-3z_2+3x^2)D_{26} \nn\\
     &\quad -x^2D_{32}-D_{34}+z_2D_{36}-(1-z_3-x^2)D_{38}
            +z_1D_{310} -2(2D_{27}+D_{312})/m_t^2\big] \nn\\
     &\quad \times m_t^4/4\pi, \nn\\
 c^{(f)}_{L2}
 =&-C_A\big[ 2z_3D_0+2(1+z_3)D_{11}+(3-4z_2+3x^2)D_{12}
            -(2+z_1-2z_2+2x^2)D_{13} \nn\\
     &\quad +5D_{21}+3x^2D_{22}-4z_2D_{24}-5z_1D_{25}
            +4(1-z_3-x^2)D_{26}+6D_{27}/m_t^2\big]\,m_t^4/8\pi, \nn\\
 c^{(f)}_{L3}
 =&-C_A\big[-2z_3(D_0+2D_{11})+(z_3-z_1)D_{12}+z_1D_{13}
            -(1+2z_3)D_{21}-3x^2D_{22} \nn\\
     &\quad +2(1-z_1+x^2)D_{24}-2(1-2z_1-z_2+x^2)D_{25}
            -(1-z_3-x^2)(3D_{26}+D_{310}) \nn\\
     &\quad -D_{31}+z_2D_{34}+z_1D_{35}-x^2D_{36}
            -2(7D_{27}+5D_{311})/m_t^2\big]\,m_t^4/4\pi, \nn\\
 c^{(f)}_{L4}
 =&-C_A\big[-2D_{11}+2(z_2-x^2)D_{12}+(1-z_2+x^2)(D_{13}+D_{23})
            -2D_{21}-2x^2D_{22} \nn\\
     &\quad +2z_2D_{24}+(2-3z_2)D_{25}+(z_2+2x^2)D_{26}
            +D_{35}-z_1D_{37}+x^2D_{38} \nn\\
     &\quad +(1-z_3-x^2)D_{39}-z_2D_{310}-2(2D_{27}-5D_{313})/m_t^2
           \big]\,m_t^4/4\pi, \nn\\
  c^{(f)}_{L6}
 =&-C_A\big[ 2D_{11}+(3-5z_2+3x^2)D_{12}-2(z_1-z_2+x^2)D_{13}
            +4D_{21}-(z_2-4x^2)D_{22} \nn\\
     &\quad +z_1D_{23}+(2-5z_2)D_{24}-2(1+2z_1-z_2)D_{25}
            -(3-2z_1-4z_2+5x^2)D_{26} \nn\\
     &\quad +x^2D_{32}+D_{34}-D_{35}-z_2D_{36}+z_1D_{37}
            +(1-z_3-2x^2)D_{38}-(1-z_3-x^2)D_{39} \nn\\
     &\quad -(z_1-z_2)D_{310}+2(3D_{27}+D_{312}-D_{313})/m_t^2
           \big]\,m_t^4/4\pi, \nn\\
 c^{(f)}_{L8}
 =&-C_A\big[2(D_{12}-D_{13}+D_{24}-D_{25})+D_{22}-D_{26}+D_{36}-D_{310}
       \big]\,m_t^4/\pi, \nn\\
 c^{(f)}_{L10}
 =&-C_A\big[D_{23}-D_{26}-D_{38}+D_{39}\big]\,m_t^4/\pi, \nn\\
 c^{(f)}_{R1}
 =&-C_A\big[ 2z_3D_0+2(1+2z_3)D_{11}-(2-2z_1+z_2-2x^2)D_{12}
            -(2+z_1-2z_2+2x^2)D_{13} \nn\\
     &\quad +5D_{21}+3x^2D_{22}-4(z_2D_{24}+z_1D_{25})
            +(1-z_3-x^2)(3D_{26}+D_{310})+D_{31} \nn\\
     &\quad -z_2D_{34}-z_1D_{35}+x^2D_{36}+2(5D_{27}+D_{311})/m_t^2
           \big]\,m_t^4/4\pi, \nn\\
 c^{(f)}_{R2}
 =&-C_A\big[-2z_3D_0
            -(11-4z_1-5z_2+x^2)D_{11}+(2-2z_1+z_2-2x^2)D_{12} \nn\\
     &\quad +(2+z_1-2z_2+2x^2)D_{13}
            -(5+z_2)D_{21}-5x^2D_{22}+(5z_2+2x^2)D_{24} \nn\\
     &\quad -(1-6z_1-z_2+x^2)D_{25}-5(1-z_3-x^2)D_{26}-12D_{27}/m_t^2
           \big]\,m_t^4/8\pi, \nn\\
 c^{(f)}_{R3}
 =&-C_A\big[-D_{11}+D_{12}-2(D_{21}-D_{24})\big]\,m_t^4/2\pi, \nn\\
 c^{(f)}_{R4}
 =&-C_A\big[D_{11}-D_{12}+2(D_{25}-D_{26})\big]\,m_t^4/2\pi, \nn\\
 c^{(f)}_{R6}
 =&-C_A\big[-D_{11}+2D_{12}-D_{13}+D_{24}-D_{25}\big]\,m_t^4/2\pi, \nn\\
 c^{(f)}_{R8}
 =&-C_A\big[-D_{11}+D_{13}-2(D_{21}-D_{25})-D_{34}+D_{35}\big]\,m_t^4/\pi,
  \nn\\
 c^{(f)}_{R10}
 =&-C_A\big[-D_{23}+2D_{25}-D_{26}-D_{37}+D_{310}\big]\,m_t^4/\pi.
\end{align}


\end{document}